\documentclass[english]{article}
\usepackage[T1]{fontenc}
\usepackage[latin9]{inputenc}
\usepackage{geometry}
\usepackage{amsthm}
\usepackage{amsmath}
\usepackage{amssymb}

\makeatletter
\theoremstyle{plain}
\newtheorem{thm}{\protect\theoremname}
  \theoremstyle{plain}
  \newtheorem{lem}[thm]{\protect\lemmaname}

\makeatother

\usepackage{babel}
  \providecommand{\lemmaname}{Lemma}
\providecommand{\theoremname}{Theorem}

\begin{document}
\global\long\def\opstyle#1{\mathbb{#1}}

\global\long\def\ex#1{\mathbb{E}\left[#1\right]}

\global\long\def\E{\mathbb{E}}

\global\long\def\sm{\setminus}

\global\long\def\a{\alpha}

\global\long\def\b{\beta}

\global\long\def\pab{p_{\a\b}}

\global\long\def\ab{\a\b}

\global\long\def\L{L_{GRD}}
\global\long\def\R{R_{GRD}}

\global\long\def\xp#1{\mathbb{E}\left[#1\right]}

\global\long\def\pr#1{\opstyle P \left[ #1 \right]}

\global\long\def\prcond#1#2{\opstyle P \left[\left. #1 \right\vert #2 \right]}

\global\long\def\excond#1#2{\opstyle E \left[\left. #1 \right\vert #2 \right]}

\global\long\def\exls#1#2{\opstyle{\opstyle E}_{#1}\left[ #2 \right]}

\global\long\def\prls#1#2{\opstyle P_{#1}\left[ #2 \right]}

\global\long\def\br#1{\left( #1 \right)}

\global\long\def\brq#1{\left[ #1 \right]}

\global\long\def\brw#1{\left\{  #1\right\}  }

\global\long\def\size#1{\left|#1\right|}

\global\long\def\setst#1#2{\left\{  #1\left|#2\right.\right\}  }

\global\long\def\setst#1#2{\left\{  \left.#1\right|#2\right\}  }

\global\long\def\setstcol#1#2{\left\{  #1:#2\right\}  }

\global\long\def\set#1{\left\{  #1\right\}  }

\global\long\def\adj#1{\mbox{\scriptsize Adj}\br{#1}}

\global\long\def\st#1{[#1] }

\global\long\def\rin#1{r^{in}\br{#1}}

\global\long\def\rout#1{r^{out}\br{#1}}

\global\long\def\rini#1#2{r_{#1}^{in}\br{#2}}

\global\long\def\routi#1#2{r_{#1}^{out}\br{#2}}

\global\long\def\Iin{{\cal I}^{in}}

\global\long\def\Iout{{\cal I}^{out}}

\global\long\def\kin{k^{in}}

\global\long\def\kout{k^{out}}

\global\long\def\Min#1{{\cal M}_{#1}^{in}}

\global\long\def\Mout#1{{\cal M}_{#1}^{out}}

\global\long\def\M{{\cal M}}

\global\long\def\deltaM{\delta^{\M}}

\global\long\def\betin#1{\beta_{#1}^{in}}

\global\long\def\betout#1{\beta_{#1}^{out}}

\global\long\def\Bin#1{B_{#1}^{in}}

\global\long\def\B#1{B_{#1}}

\global\long\def\Bout#1{B_{#1}^{out}}

\global\long\def\indi#1{\chi\brq{#1}}

\global\long\def\chr#1{\mathbf{1}_{#1}}

\global\long\def\pxE{\br{p_{e}x_{e}}_{e\in E}}

\global\long\def\pxEt#1{\br{p_{e}x_{e}}_{e\in E^{#1}}}

\global\long\def\xE{\br{x_{e}}_{e\in E}}

\global\long\def\xEt#1{\br{x_{e}}_{e\in E^{#1}}}

\global\long\def\e{\bar{e}}

\global\long\def\evalat#1#2{ #1 \Big|_{#2}}

\global\long\def\Df#1#2{\frac{\partial#1}{\partial#2}}

\global\long\def\event#1{\mathcal{E}_{#1}^{out}}
 \global\long\def\eventsucc#1{\mathcal{E}_{#1}^{in}}

\title{Greedy algorithm for stochastic matching is a 2-approximation}

\author{Marek Adamczyk \thanks{Department of Computer, Control, and Management Engineering, Sapienza University of Rome, Italy; e-mail: \texttt{adamczyk@dis.uniroma1.it}}
}
\maketitle
\begin{abstract}
{\normalsize Motivated by applications in online dating and kidney
exchange, the stochastic matching problem was introduced by Chen,
Immorlica, Karlin, Mahdian and Rudra~(2009). They have proven a 4-approximation
of a simple greedy strategy, but conjectured that it is in fact a
2-approximation. In this paper we confirm this hypothesis.}{\normalsize \par}
\end{abstract}

\section{{\normalsize Introduction}}

We are given an undirected graph $G=(V,E)$ in which every edge $uv\in E$
is assigned a real number $p_{uv}\in\brq{0,1}$. Every vertex $v\in V$
is assigned a positive integer number $t_{v}$, called \textit{patience
}. At every step we can \textit{probe} any edge $uv\in E$, but only
if $t_{u}>0$ and $t_{v}>0$. Probe of $uv$ edge will end up with
success with probability $p_{uv}$. In this case vertices $u$ and
$v$ will be removed from the graph, as well as all the edges incident
to $u$ and $v$. With probability $1-p_{uv}$ the probe fails. In
this case edge $uv$ is removed from the graph, and patience numbers
$t_{u}\mbox{ and }t_{v}$ are both decreased by $1$. If after a certain
step, patience $t_{v}$ of a vertex $v$ becomes $0$, then we remove
vertex $v$ from the graph, together with all edges incident to it.
The outcome of a strategy is the total number of edges successfully
probed, and our goal is to maximize the average outcome of a strategy.

\subsection{{\normalsize Motivation }\cite{DBLP:conf/icalp/ChenIKMR09}}

\paragraph{Kidney Exchange}

A patient awaiting a kidney transplant can receive the organ from
a living friend or a family member. Unfortunately, even if someone
is willing to donate a kidney to a patient, it may happen that the
donor is incompatible. However, it is possible to find two such incompatible
patient/donor pairs where each donor is compatible with the patient
from other pair. Four operations are then performed simultaneously
resulting in two kidney transplants. In year 2000 the United Network
for Organ Sharing (UNOS) launched a program of such kidney exchanges.

In order to maximize the number of transplanted kidneys we need to
find the maximum matching in a graph, nodes of which represent incompatible
patient/donor pairs. However, this graph is not given entirely upfront.
For two incompatible pairs we need to run three tests to find out
if we can perform a kidney exchange between them. First two tests
are ``easy'', and estimate the probability that the third ``hard''
test will be successful --- three passed tests allow to perform a
kidney exchange. Medical characteristics of the program, imply that
the third test cannot be performed for every two patient/donor pairs.
Also, the transplants have to be performed immediately after we have
matched two pairs. Thus we can see that the graph of patient/donor
pairs is in fact random. Moreover, we also need to model the fact
that a patient has limited time for awaiting a kidney.

\paragraph{Online Dating}

Consider an online dating service. For each pair of users, machine
learning algorithms can estimate the probability that the given pair
will form a happy couple. However, only after a pair meets we know
for sure if they were successfully matched (and together left the
dating service). Users have individual patience numbers that bound
how many unsuccessful dates they are willing to go on until they will
leave the dating service forever. The objective of the service is
to maximize the number of successfully matched couples.

To model this as a stochastic matching problem, users are represented
as vertices $V$ of a graph $G=\br{V,E}$. Every edge $uv\in E$ corresponds
to a date between users $u$ and $v$, with $p_{uv}$ being the probability
that a couple $u,v$ forms a happy couple after a date. Successfully
probed edges have to form a matching --- a user can be in at most
one couple. If we assume that $u$ is willing to go for at most $t\br u$
unsuccessful dates, then we can probe at most $t\br u$ edges adjacent
to user $u$.

\subsection{Related work}

The stochastic matching problem together with applications to online
dating and kidney exchange was introduced by Chen~et~al.\ \cite{DBLP:conf/icalp/ChenIKMR09},
where authors proved a 4-approximation of a greedy strategy for unweighted
case. They also proposed weighted variant of the problem, and showed
that simple greedy rules can be arbitrary bad in this case. First
constant approximation for the weighted case was given by Bansal~et~al.\ \cite{Bansal:woes}
who gave 3-approximation for bipartite graphs and 4-approximation
for non-bipartite graphs.

The stochastic matching problem falls into the class of adaptive stochastic
optimization problems. Dean~et~al.\ \cite{DBLP:journals/mor/DeanGV08}
were first to consider adaptivity in stochastic optimization. In this
class of problems, the solution is in fact a process, and the optimal
one might even require larger than polynomial space to describe. Since
the work of Dean~et~al.\ a number of such problems were introduced~\cite{DBLP:conf/soda/GuhaM07,DBLP:conf/stoc/GuhaM07,DBLP:conf/wine/AsadpourNS08,DBLP:conf/latin/GoemansV06,DBLP:conf/soda/DeanGV05}.

\section{{\normalsize Preliminaries}}

If at a certain step an algorithm probes edge coming out of vertex
$\alpha\in G$, then we say that the algorithm \textit{probes vertex
$\alpha$}. We shall also say that $v$ is \emph{taken into the matching}
if one of edges incident to $v$ will be successfully probed.

We call an algorithm \textit{deterministic}, if in each step the choice
of an edge to probe is unambiguous and depends only on previous steps.
We call an algorithm \emph{randomized} when at least once the choice
of an edge to probe is random.

If $ALG$ denotes an algorithm, then $\mathbb{E}ALG$ is the expected
number of edges taken into the matching by this algorithm. 

The instance of our problem is a pair $\br{G,t}$, where $G$ is a
random graph with given probabilities of edges, and $t:V\mapsto\mathbb{Z}_{\geq0}$
is the patience function. For an algorithm $ALG$ we will denote by
$\br{G^{ALG},t^{ALG}}$ the instance of the problem on which $ALG$
is executed. We call an instance $\br{G',t'}$ a \emph{sub-instance}
of the instance $\br{G,t}$, if $G'$ is a proper subgraph of $G$
and $t'_{v}\leq t_{v}$ for every vertex $v\in G'$.

It is reasonable to consider also an instance formed by the empty
graph, because it may appear in the inductive reasoning. Of course,
in this case performance of any algorithm is zero.

The optimal algorithm on the instance $(G,t)$ is denoted by $OPT\br G$
--- we do not write $OPT(G,t)$, because it is always clear from the
context which patience numbers we are using. We can assume without
loss of generality that $OPT(G)$ is deterministic. 

Greedy algorithm probes at each step an edge with the maximum probability
of success; ties broken arbitrarily. Greedy algorithm on graph $G$
is denoted by $GRD\br G$ --- since this algorithm does not consider
patience numbers, we do not write $GRD\br{G,t}$.

\paragraph{Decision tree}

Each deterministic algorithm $ALG$ can be represented by an (exponential-sized)
\emph{decision tree} $T_{ALG}$. Each node of that tree corresponds
to a probe of an edge. Let $\ab$ be the edge that the algorithm $ALG$
probes first. The root $r\in T_{ALG}$ represents the probe of $\ab$,
and has assigned value $p_{r}$, that is equal to $\pab$. The left
subtree of the root $r$ represents the proceeding of the algorithm
$ALG$ after a successful probe of the edge $\ab$, and the right
subtree --- after a failure. More precisely: 
\begin{itemize}
\item the left subtree corresponds to the algorithm on the instance $\br{G\sm\brw{\a,\b},t}$;
\item the right subtree corresponds to the algorithm on the instance $\br{G\sm\brw{\ab},t'}$,
where $t'_{\a}=t_{\a}-1,t'_{\b}=t_{\b}-1$ and $t'_{\gamma}=t_{\gamma}$
for other vertices $\gamma$.
\end{itemize}
Let us notice that in the first point we remove both vertices $\a$
and $\b$ from the graph, while in the second point we only remove
the edge $\ab$. The definition of the left and right subtree of $r$
is recursive.

The probability of reaching a node $v\in T_{ALG}$ will be denoted
by $q_{v}$. If the left edge of a node $v$ was labeled $p_{v}$
and the right edge by $1-p_{v}$, then $q_{v}$ would be the product
of labels on all edges on the path from the root of $T_{ALG}$ to
the node $v$. For a tree $T$ we denote the sum $\sum_{v\in T}q_{v}p_{v}$
by $\E T$. The performance of an algorithm $ALG$ can be expressed
using the decision tree: 
\[
\E ALG=\E T_{ALG}=\sum_{v\in T_{ALG}}q_{v}p_{v}.
\]
 For a node $v\in T_{ALG}$, we denote by $T\br v$ the subtree of
$T_{ALG}$ rooted at $v$, and by $L\br v$, $R\br v$ its left and
right subtree respectively.

Throughout the paper we distinguish between nodes and vertices, i.e.\
nodes only belong to a decision tree, and vertices only belong to
a random graph. Moreover, vertices and only them are denoted by Greek
letters.

Since we assume that the optimal algorithm is deterministic, we can
represent it by such a decision tree. We also assume that every subtree
of the tree $T_{OPT}$ representing optimal algorithm is optimal on
its instance, even when probability of reaching such subtree is zero.

\section{{\normalsize Analysis of the greedy algorithm}}

This whole section is a proof of the following theorem.
\begin{thm}
\label{theorem} Greedy algorithm is a $2$-approximation for any
instance $\br{G,t}$ of the stochastic matching problem. 
\end{thm}
We will start with a lemma from \cite{DBLP:conf/icalp/ChenIKMR09}.
\begin{lem}
\label{lemma1} For any node $v\in T_{OPT}$, $\mathbb{E}T\br v\leq\mathbb{E}L\br v+1$.\end{lem}
\begin{proof}
Algorithm which follows $R(v)$ is a proper algorithm for an instance
on which $T(v)$ works, so $\mathbb{E}R\br v\leq\mathbb{E}T\br v$
since every subtree of $T_{OPT}$ is optimal. Hence 
\[
\mathbb{E}T\br v=p_{v}\br{1+\mathbb{E}L\br v}+(1-p_{v})\mathbb{E}R\br v\leq p_{v}\br{1+\mathbb{E}L\br v}+\br{1-p_{v}}\mathbb{E}T\br v.
\]
This gives $p_{e}\mathbb{E}T\br v\leq p_{e}\br{1+\mathbb{E}L\br v}$,
and finally $\mathbb{E}T\br v\leq1+\mathbb{E}L\br v$.
\end{proof}
The proof is inductive with respect to sub-instances of problem $\br{G,t}$.
The case when the graph has no edges is trivial, so suppose it has
at least one edge. Let $\alpha\beta$ be the first edge probed by
the greedy algorithm. Let $T_{GRD}$ be the decision tree of the greedy
algorithm. We denote by $L_{GRD}$ the algorithm that follows the
left subtree of the tree $T_{GRD}$ --- it represents the execution
of $GRD$ after the successful probe of $\ab$. Analogically, $R_{GRD}$
denotes the algorithm that follows the right subtree. According to
the notation, the instance on which $\L$ works is denoted by $\br{G^{\L},t^{\L}}$
or shortly $G^{\L}$; analogically, $G^{\R}$ is the instance on which
$\R$ works. Note that $\E GRD\br G=\pab+\pab\cdot\E\L+\br{1-p_{\alpha\beta}}\cdot\E\R$

When it does not make a problem, we use $OPT$ instead of $OPT\br G$.

\subsection{{\normalsize Algorithm for the instance $G^{\L}$}}

Let $X$ be the set of nodes of $T_{OPT}$ which correspond to the
probe of edge $\alpha\beta$. Let us define an algorithm $OPT'$ that
follows algorithm $OPT\br G$ until it reaches a node $x\in X$. After
reaching node $x$, $OPT'$ probes edge $\alpha\beta$, but afterwards
goes straight to the subtree $L\br x$, regardless of the probe result.
In other words, after the probe it behaves like if the probe was successful,
even if it was not. In Appendix we comment on the necessity of using
such a modification of $OPT$ in the proof.
\begin{lem}
\label{lem:optprim}For algorithm $OPT'$ defined as above it holds
that 
\[
\mathbb{E}OPT\leq\mathbb{E}OPT'+(1-p_{\alpha\beta})\mathbb{P}(OPT\mbox{ \emph{probes} }\alpha\beta).
\]
\end{lem}
\begin{proof}
For a node $x\in X$ let $T\br x$ denote the subtree of $T_{OPT}$
rooted at $x$; we denote also $T\br X=\bigcup_{x\in X}T\br x$. Now
we have: 
\begin{align*}
\mathbb{E}OPT=\E T_{OPT} & =\sum_{v\in T_{OPT}\setminus T(X)}q_{v}p_{v}+\sum_{x\in X}q_{x}\mathbb{E}T(x).\\
\intertext{\mbox{From Lemma\,\ref{lemma1} we get that}}\mathbb{E}OPT & \leq\sum_{v\in T_{OPT}\setminus T(X)}q_{v}p_{v}+\sum_{x\in X}q_{x}(1+\mathbb{E}L(x)).\\
\intertext{\mbox{On the other hande, the average outcome of \ensuremath{OPT'} is equal to}}\mathbb{E}OPT' & =\sum_{v\in T_{OPT}\setminus T(X)}q_{v}p_{v}+\sum_{x\in X}q_{x}(p_{\alpha\beta}+\mathbb{E}L(x)).\\
\intertext{\mbox{Hence}}\mathbb{E}OPT & \leq\sum_{v\in T_{OPT}\setminus T(X)}q_{v}p_{v}+\sum_{x\in X}q_{x}(1+\mathbb{E}L(x))\\
 & =\sum_{v\in T_{OPT}\setminus T(X)}q_{v}p_{v}+\sum_{x\in X}q_{x}(p_{\alpha\beta}+\mathbb{E}L(x))+\sum_{x\in X}q_{x}(1-p_{\alpha\beta})\\
 & =\mathbb{E}OPT'+(1-p_{\alpha\beta})\sum_{x\in X}q_{x}.
\end{align*}
It remains to notice that $\sum\limits _{x\in X}q_{x}=\mathbb{P}(OPT\mbox{ probes }\alpha\beta)$. 
\end{proof}
Now we use algorithm $OPT'$ to construct an algorithm for the instance
$(G^{L_{GRD}},t^{L_{GRD}})$. Suppose we are given a black-box that
executes the algorithm $OPT'$:
\begin{itemize}
\item we give to the black-box the initial instance $\br{G,t}$;
\item the black-box outputs us the first edge to probe;
\item we make this probe;
\item we give back the result of this probe to the black-box; 
\item the black-box outputs a second edge to probe, we probe the second
edge, we give back the result, and so on.
\end{itemize}
Given this black-box we can define algorithm $ALG_{L}$ for the instance
$(G^{L_{GRD}},t^{L_{GRD}})$:
\begin{itemize}
\item We give to the black-box the initial instance $\br{G,t}$.
\item When the black-box outputs an edge to be probed, which is \textbf{not}
adjacent to $\a$ nor $\b$, then $ALG_{L}$ makes that probe.
\item When the black-box outputs an edge $e$ to be probed, which is adjacent
to $\a$ or $\b$, then the algorithm $ALG_{L}$ \emph{fakes} the
probe of $e$:

\begin{itemize}
\item a fake probe means that $ALG_{L}$ makes only a coin toss, and reports
a result of a probe, but does not probe any edge;
\item the coin toss is distributed according to the probability of an edge,
i.e.\ with probability $p_{e}$ it tells the black-box that the probe
succeeded, and with probability $1-p_{e}$ it tells the black-box
that the probe failed.
\end{itemize}
\end{itemize}
Another definition can be given that uses decision trees. $ALG_{L}$
follows the decision tree of $OPT'$, but when it reaches a node $v$,
which corresponds to a probe of vertex $\alpha$ or $\beta$, then
$ALG_{L}$ flips a coin, and with probability $p_{v}$ it goes to
the left subtree $L\br v$, and with probability $1-p_{v}$ it goes
to the right subtree $R\br v$, and no actual probe is made then.

This (randomized) algorithm is a feasible algorithm for the instance
$(G^{L_{GRD}},t^{L_{GRD}})$, because graph $G^{L_{GRD}}$ is made
from $G$ by removing vertices $\alpha$ and $\beta$, and all edges
adjacent to them. Moreover, for every vertex $v\in G^{L_{GRD}}$ we
have $t_{v}^{L_{GRD}}=t_{v}$. Performance of algorithm $ALG_{L}$
is equal to the performance of $OPT'$ minus penalty for skipped probes.
Let us denote this penalty by $S_{L}$. From the definition $\mathbb{E}OPT'=\mathbb{E}ALG_{L}+\mathbb{E}S_{L}.$

Let us analyze $\E S_{L}$ more carefully.
\begin{itemize}
\item If $OPT'$ probes edge $\alpha\beta$, then:

\begin{itemize}
\item with probability $p_{\ab}$ it succeeds, and afterwards all probes
of $OPT'$ do not probe $\alpha$ nor $\beta$;
\item with probability $1-\pab$ it fails, but because of the definition
of $OPT'$, after the failed probe of $\ab$, $OPT'$ behaves like
if the probe was successful, so, also in this case, $OPT'$ does not
probe $\a$ nor $\b$ afterwards. 
\end{itemize}
\item If $OPT'$ does not probe $\alpha\beta$, then the penalty is just
equal to the conditional expected number of successfully probed edges
adjacent to $\alpha\beta$: 
\[
\prcond{OPT'\mbox{ takes }\alpha}{OPT'\mbox{ does not probe }\alpha\beta}+\prcond{OPT'\mbox{ takes }\b}{OPT'\mbox{ does not probe }\alpha\beta}.
\]

\end{itemize}
Thus we can write that $\mathbb{E}S_{L}$ is equal to 
\begin{align*}
\pr{OPT'\mbox{ probes }\alpha\beta}p_{\alpha\beta}+\pr{OPT'\mbox{ does not probe }\alpha\beta}\Bigl( & \prcond{OPT'\mbox{ takes }\alpha}{OPT'\mbox{ does not probe }\alpha\beta}\\
+ & \prcond{OPT'\mbox{ takes }\beta}{OPT'\mbox{ does not probe }\alpha\beta}\Bigr).
\end{align*}
From the definition, $OPT'$ works just like $OPT$, unless it reaches
$\alpha\beta$, so the above expression is in fact equal to 
\begin{align*}
\pr{OPT\mbox{ probes }\alpha\beta}p_{\alpha\beta}+\pr{OPT\mbox{ does not probe }\alpha\beta}\Bigl( & \prcond{OPT\mbox{ takes }\alpha}{OPT\mbox{ does not probe }\alpha\beta}\\
+ & \prcond{OPT\mbox{ takes }\beta}{OPT\mbox{ does not probe }\alpha\beta}\Bigr).
\end{align*}
Let us introduce a shorter notation. Denote the event that $OPT$
probes $\alpha\beta$ as ``probe $\alpha\beta$'', and ``$\neg\mbox{probe }\alpha\beta$''
as the opposite event. The event that $OPT$ takes $\alpha$ under
the condition that the edge $\alpha\beta$ is not probed, is denoted
as ``$\mbox{take }\alpha|\neg\mbox{probe }\alpha\beta$''; analogically
for $\b$. Now we can write that 
\[
\mathbb{E}S_{L}=\pr{\mbox{probe }\alpha\beta}p_{\alpha\beta}+\pr{\neg\mbox{probe }\alpha\beta}\Bigl(\prcond{\mbox{take }\alpha}{\neg\mbox{probe }\alpha\beta}+\prcond{\mbox{take }\mbox{\ensuremath{\b}}}{\neg\mbox{probe }\alpha\beta}\Bigr).
\]
We join the inequality from Lemma~\ref{lem:optprim} with the above
equality, and we get that 
\begin{eqnarray}
\mathbb{E}OPT & \leq & \mathbb{E}OPT'+(1-p_{\alpha\beta})\pr{\mbox{probe }\alpha\beta}\nonumber \\
 & = & \mathbb{E}ALG_{L}+\mathbb{E}S_{L}+(1-p_{\alpha\beta})\pr{\mbox{probe }\alpha\beta}\nonumber \\
 & = & \mathbb{E}ALG_{L}+\br{1-p_{\alpha\beta}}\pr{\mbox{probe }\alpha\beta}+\pr{\mbox{probe }\alpha\beta}p_{\alpha\beta}\nonumber \\
 &  & +\pr{\neg\mbox{probe }\alpha\beta}\Bigl(\prcond{\mbox{take }\alpha}{\neg\mbox{probe }\alpha\beta}+\prcond{\mbox{take }\mbox{\ensuremath{\b}}}{\neg\mbox{probe }\alpha\beta}\Bigr)\nonumber \\
 & = & \mathbb{E}ALG_{L}+\pr{\mbox{probe }\alpha\beta}+\pr{\neg\mbox{probe }\alpha\beta}\Bigl(\prcond{\mbox{take }\alpha}{\neg\mbox{probe }\alpha\beta}+\prcond{\mbox{take }\mbox{\ensuremath{\b}}}{\neg\mbox{probe }\alpha\beta}\Bigr).\label{eq:algl}
\end{eqnarray}

\subsection{{\normalsize Algorithm for the instance $G^{\R}$}}

The instance $\br{G^{R_{GRD}},t^{R_{GRD}}}$ is made of $\br{G,t}$
by removing the edge $\alpha\beta$ and decreasing the patience of
$\alpha$ and $\beta$, i.e.\ $t_{\alpha}^{R_{GRD}}=t_{\alpha}-1$
and $t_{\beta}^{R_{GRD}}=t_{\beta}-1$. To define algorithm for $G^{\R}$
we use the same type of black-box definition. Suppose we are given
a black-box that executes the $OPT$. 
\begin{itemize}
\item We give to the black-box the initial instance $\br{G,t}$.
\item When the black-box outputs an edge $\ab$ to be probed, then we fake
the probe --- with probability $\pab$ we say it was a success, with
$1-\pab$ we say it was a failure, although we do not probe any edge
at all.
\item When the black-box outputs an edge to be probed, and it would be the
probe number $t_{\a}$ of vertex $\a$ made by algorithm $ALG_{R}$,
then we fake the probe; similarly with $\beta$.
\end{itemize}
We need to notice a crucial feature of this construction. Suppose
that $OPT$ attempts to probe $\ab$, and $ALG_{R}$ fakes the probe
then. Thanks to this omitted probe, $ALG_{R}$ saves one unit of patience
for both $\a$ and $\b$. It means that even if later $OPT$ makes
a probe number $t_{\a}$ of vertex $\a$, then this is actually a
probe number $t_{\a}-1$ for $ALG_{R}$, and $ALG_{R}$ can make this
probe. Hence, if $OPT$ probes $\ab$, then the fake probe of $\ab$
is the only probe that $ALG_{R}$ fakes in the whole execution of
$OPT$ .

Let $S_{R}$ be the penalty for the faked probes. Similarly as before
we can write that $\mathbb{E}OPT=\mathbb{E}ALG_{R}+\mathbb{E}S_{R}.$
From the definition of the faked probes it follows that 
\begin{eqnarray*}
\mathbb{E}S_{R} & = & \pr{OPT\mbox{ probes }\alpha\beta}p_{\alpha\beta}\\
 &  & +\pr{OPT\mbox{ does not probe }\alpha\beta\mbox{ and takes }\alpha\mbox{ in probe number }t_{\alpha}}\\
 &  & +\pr{OPT\mbox{ does not probe }\alpha\beta\mbox{ and takes }\mbox{\ensuremath{\b}}\mbox{ in probe number }t_{\b}}.
\end{eqnarray*}
Let us introduce shorter notation also this time. Instead of $``OPT\mbox{ takes }\alpha\mbox{ in probe number }t_{\alpha}$''
we shall write ``$\mbox{take }\alpha\mbox{ in }t_{\alpha}$''; analogically
with $\beta$. Now we can write the above equality shorter: 
\begin{eqnarray*}
\mathbb{E}S_{R} & = & \pr{\mbox{probe }\alpha\beta}p_{\alpha\beta}+\pr{\neg\mbox{probe }\alpha\beta\wedge\mbox{take }\alpha\mbox{ in }t_{\alpha}}+\pr{\neg\mbox{probe }\alpha\beta\wedge\mbox{take }\beta\mbox{ in }t_{\beta}}\\
 & = & \pr{\mbox{probe }\alpha\beta}p_{\alpha\beta}+\pr{\neg\mbox{probe }\alpha\beta}\Bigl(\prcond{\mbox{take }\alpha\mbox{ in }t_{\alpha}}{\neg\mbox{probe }\alpha\beta}+\prcond{\mbox{take }\beta\mbox{ in }t_{\beta}}{\neg\mbox{probe }\alpha\beta}\Bigr).
\end{eqnarray*}
Finally we can write that 
\begin{eqnarray}
\mathbb{E}OPT & = & \mathbb{E}ALG_{R}+\E S_{R}\nonumber \\
 & = & \E ALG_{R}+\pr{\mbox{probe }\alpha\beta}p_{\alpha\beta}\nonumber \\
 &  & +\pr{\neg\mbox{probe }\alpha\beta}\Bigl(\prcond{\mbox{take }\alpha\mbox{ in }t_{\alpha}}{\neg\mbox{probe }\alpha\beta}+\prcond{\mbox{take }\beta\mbox{ in }t_{\beta}}{\neg\mbox{probe }\alpha\beta}\Bigr).\label{eq:algr}
\end{eqnarray}

\subsection{{\normalsize Combining $ALG_{L}$ and $ALG_{R}$}}

We multiply inequality~(\ref{eq:algl}) by $p_{\alpha\beta}$, we
multiply equality~(\ref{eq:algr}) by $1-p_{\alpha\beta}$, we add
them together, and we obtain:
\begin{eqnarray*}
\E OPT & \leq & \pab\left\{ \mathbb{E}ALG_{L}+\pr{\mbox{probe }\alpha\beta}+\pr{\neg\mbox{probe }\alpha\beta}\Bigl(\prcond{\mbox{take }\alpha}{\neg\mbox{probe }\alpha\beta}+\prcond{\mbox{take }\mbox{\ensuremath{\b}}}{\neg\mbox{probe }\alpha\beta}\Bigr)\right\} \\
 &  & +\br{1-\pab}\left\{ \E ALG_{R}+\pr{\mbox{probe }\alpha\beta}p_{\alpha\beta}\right\} \\
 &  & +\br{1-\pab}\left\{ \pr{\neg\mbox{probe }\alpha\beta}\Bigl(\prcond{\mbox{take }\alpha\mbox{ in }t_{\alpha}}{\neg\mbox{probe }\alpha\beta}+\prcond{\mbox{take }\beta\mbox{ in }t_{\beta}}{\neg\mbox{probe }\alpha\beta}\Bigr)\right\} .
\end{eqnarray*}
After grouping terms in the above expression we get
\begin{eqnarray*}
\E OPT & \leq & p_{\alpha\beta}\mathbb{E}ALG_{L}+\br{1-p_{\alpha\beta}}\mathbb{E}ALG_{R}\\
 & + & p_{\alpha\beta}\pr{\mbox{probe }\alpha\beta}\cdot\br{2-p_{\alpha\beta}}\\
 & + & p_{\alpha\beta}\pr{\neg\mbox{probe }\alpha\beta}\br{\prcond{\mbox{take }\alpha}{\neg\mbox{probe }\alpha\beta}+\frac{1-p_{\alpha\beta}}{p_{\alpha\beta}}\prcond{\mbox{take }\alpha\mbox{ in }t_{\alpha}}{\neg\mbox{probe }\alpha\beta}}\\
 & + & p_{\alpha\beta}\pr{\neg\mbox{probe }\alpha\beta}\left(\prcond{\mbox{take }\b}{\neg\mbox{probe }\alpha\beta}+\frac{1-p_{\alpha\beta}}{p_{\alpha\beta}}\prcond{\mbox{take }\b\mbox{ in }t_{\b}}{\neg\mbox{probe }\alpha\beta}\right).
\end{eqnarray*}
To finish the proof it remains to show that 
\begin{equation}
\prcond{\mbox{take }\alpha}{\neg\mbox{probe }\alpha\beta}+\frac{1-p_{\alpha\beta}}{p_{\alpha\beta}}\prcond{\mbox{take }\alpha\mbox{ in }t_{\alpha}}{\neg\mbox{probe }\alpha\beta}\leq1.\label{eq:exchangeineq}
\end{equation}
This inequality, and analogical for $\b$, will imply that 
\begin{eqnarray*}
\E OPT & \leq & p_{\alpha\beta}\mathbb{E}ALG_{L}+\br{1-p_{\alpha\beta}}\mathbb{E}ALG_{R}+p_{\alpha\beta}\pr{\mbox{probe }\alpha\beta}\cdot\br{2-p_{\alpha\beta}}+p_{\alpha\beta}\pr{\neg\mbox{probe }\alpha\beta}\cdot2\\
 & \leq & p_{\alpha\beta}\mathbb{E}ALG_{L}+\br{1-p_{\alpha\beta}}\mathbb{E}ALG_{R}+2p_{\alpha\beta}\br{\pr{\mbox{probe }\alpha\beta}+\pr{\neg\mbox{probe }\alpha\beta}}\\
 & = & p_{\alpha\beta}\mathbb{E}ALG_{L}+\br{1-p_{\alpha\beta}}\mathbb{E}ALG_{R}+2p_{\alpha\beta}\\
 & \leq & p_{\alpha\beta}\cdot2\E\L+\br{1-p_{\alpha\beta}}\cdot2\E\R+2\pab\\
 & = & 2\E GRD\br G,
\end{eqnarray*}
where the last inequality follows from the inductive assumption.

To prove inequality~(\ref{eq:exchangeineq}) we need the following
Lemma.
\begin{lem}
\label{lem:exchange} Given that $\pab$ is the greatest probability
it holds that \emph{
\[
\frac{1-p_{\alpha\beta}}{p_{\alpha\beta}}\prcond{\mbox{take }\alpha\mbox{ in }t_{\alpha}}{\neg\mbox{probe }\alpha\beta}\leq\prcond{OPT\mbox{ does not take }\alpha\mbox{ despite of }t_{\alpha}\mbox{ probes}}{\neg\mbox{probe }\alpha\beta}.
\]
} \end{lem}
\begin{proof}
To take $\alpha$ into the matching, exactly one edge incident to
$\a$ has to be taken, so 
\[
\frac{1-p_{\alpha\beta}}{p_{\alpha\beta}}\prcond{\mbox{take }\alpha\mbox{ in }t_{\alpha}}{\neg\mbox{probe }\alpha\beta}=\frac{1-p_{\alpha\beta}}{p_{\alpha\beta}}\sum_{\gamma\in\adj{\a}}\prcond{OPT\mbox{ takes }\alpha\gamma\mbox{ in probe number }t_{\alpha}}{\neg\mbox{probe }\alpha\beta}.
\]
Edge $\alpha\gamma$ is taken into the matching, if this edge is probed
and the probe is successful, i.e.\ 
\begin{align*}
 & \frac{1-p_{\alpha\beta}}{p_{\alpha\beta}}\sum_{\gamma\in\adj{\a}}\prcond{OPT\mbox{ takes }\alpha\gamma\mbox{ in probe number }t_{\alpha}}{\neg\mbox{probe }\alpha\beta}\\
= & \frac{1-p_{\alpha\beta}}{p_{\alpha\beta}}\sum_{\gamma\in\adj{\a}}\prcond{OPT\mbox{ probes }\alpha\gamma\mbox{ in probe number }t_{\alpha}\mbox{ AND probe is successful}}{\neg\mbox{probe }\alpha\beta}.
\end{align*}
Probe number $t_{\alpha}$ is the last probe of vertex $\alpha$ regardless
of its result. Thus the result of this probe and the event that $OPT$
does not probe $\alpha\beta$ are independent. Hence 
\begin{eqnarray}
 &  & \frac{1-p_{\alpha\beta}}{p_{\alpha\beta}}\sum_{\gamma\in\adj{\a}}\prcond{OPT\mbox{ probes }\alpha\gamma\mbox{ in probe number }t_{\alpha}\mbox{ AND probe is successful}}{\neg\mbox{probe }\alpha\beta}\nonumber \\
 & = & \frac{1-p_{\alpha\beta}}{p_{\alpha\beta}}\sum_{\gamma\in\adj{\a}}\prcond{OPT\mbox{ probes }\alpha\gamma\mbox{ in probe number }t_{\alpha}}{\neg\mbox{probe }\alpha\beta}\cdot p_{\alpha\gamma}.\label{eq:indsucc}
\end{eqnarray}
Function $\frac{1-x}{x}$ is decreasing and $\pab$ is the greatest
probability, so we get 
\begin{align*}
 & \sum_{\gamma\in\adj{\a}}\frac{1-p_{\alpha\beta}}{p_{\alpha\beta}}\cdot p_{\alpha\gamma}\prcond{OPT\mbox{ probes }\alpha\gamma\mbox{ in probe number }t_{\alpha}}{\neg\mbox{probe }\alpha\beta}\\
\leq & \sum_{\gamma\in\adj{\a}}\frac{1-p_{\alpha\gamma}}{p_{\alpha\gamma}}\cdot p_{\alpha\gamma}\prcond{OPT\mbox{ probes }\alpha\gamma\mbox{ in probe number }t_{\alpha}}{\neg\mbox{probe }\alpha\beta}\\
= & \sum_{\gamma\in\adj{\a}}(1-p_{\alpha\gamma})\cdot\prcond{OPT\mbox{ probes }\alpha\gamma\mbox{ in probe number }t_{\alpha}}{\neg\mbox{probe }\alpha\beta}\\
= & \sum_{\gamma\in\adj{\a}}\prcond{OPT\mbox{ probes }\alpha\gamma\mbox{ in probe number }t_{\alpha}\mbox{ AND probe failed}}{\neg\mbox{probe }\alpha\beta}.
\end{align*}
The last equality we justify in the same way we did~(\ref{eq:indsucc}).
It remains to note that 
\begin{multline*}
\sum_{\gamma\in\adj{\a}}\prcond{OPT\mbox{ probes }\alpha\gamma\mbox{ in probe number }t_{\alpha}\mbox{ AND probe failed}}{\neg\mbox{probe }\alpha\beta}\\
=\prcond{OPT\mbox{ does not take }\alpha\mbox{ despite of }t_{\alpha}\mbox{ probes}}{\neg\mbox{probe }\alpha\beta}.
\end{multline*}
and the lemma is proved.
\end{proof}
The following sequence of inequalities proves inequality~(\ref{eq:exchangeineq}),
and therefore concludes the proof of the Theorem~1:

\begin{eqnarray*}
 &  & \prcond{\mbox{take }\alpha}{\neg\mbox{probe }\alpha\beta}+\frac{1-p_{\alpha\beta}}{p_{\alpha\beta}}\prcond{\mbox{take }\alpha\mbox{ in }t_{\alpha}}{\neg\mbox{probe }\alpha\beta}\\
 & \leq & \prcond{\mbox{take }\alpha}{\neg\mbox{probe }\alpha\beta}+\prcond{OPT\mbox{ does not take }\alpha\mbox{ despite of }t_{\alpha}\mbox{ probes}}{\neg\mbox{probe }\alpha\beta}\qquad\mbox{(from Lemma\,\ref{lem:exchange})}\\
 & \leq & \prcond{\mbox{take }\alpha}{\neg\mbox{probe }\alpha\beta}+\prcond{OPT\mbox{ does not take }\alpha}{\neg\mbox{probe }\alpha\beta}=1.
\end{eqnarray*}

\bibliographystyle{plain}
\bibliography{iterativemethod}

\section*{Appendix: why we use $OPT'$}

Recall that when we constructed $ALG_{L}$ from the algorithm $OPT'$,
then the penalty $\E S_{L}$ for faked probes could be expressed as
follows:
\begin{align*}
\E S_{L}=\pr{OPT\mbox{ probes }\alpha\beta}p_{\alpha\beta}+\pr{OPT\mbox{ does not probe }\alpha\beta}\Bigl( & \prcond{OPT\mbox{ takes }\alpha}{OPT\mbox{ does not probe }\alpha\beta}\\
+ & \prcond{OPT\mbox{ takes }\beta}{OPT\mbox{ does not probe }\alpha\beta}\Bigr).
\end{align*}
If we would make the same construction of $ALG_{L}$ but using $OPT$
instead of modified $OPT'$, then we would end up with $\E S_{L}$
equal to 
\begin{align*}
\pr{OPT\mbox{ probes }\alpha\beta}\biggl(\pab+\br{1-\pab}\Bigl( & \prcond{OPT\mbox{ takes }\alpha}{OPT\mbox{ probes }\alpha\beta\mbox{ and fails}}\\
+ & \prcond{OPT\mbox{ takes }\beta}{OPT\mbox{ probes }\alpha\beta\mbox{ and fails}}\Bigr)\biggr)\\
+\pr{OPT\mbox{ does not probe }\alpha\beta}\Bigl( & \prcond{OPT\mbox{ takes }\alpha}{OPT\mbox{ does not probe }\alpha\beta}\\
+ & \prcond{OPT\mbox{ takes }\beta}{OPT\mbox{ does not probe }\alpha\beta}\Bigr).
\end{align*}
To conclude the proof from here we would have to show that 
\[
\prcond{OPT\mbox{ takes }\alpha}{OPT\mbox{ probes }\alpha\beta\mbox{ and fails}}+\prcond{OPT\mbox{ takes }\beta}{OPT\mbox{ probes }\alpha\beta\mbox{ and fails}}\leq1.
\]
However, we don't know how to prove this inequality, and if it is
in fact true.
\end{document}